\documentstyle[preprint,aps]{revtex}     
\renewcommand{\baselinestretch}{1.346}   
\begin{document}
\draft

\title{
On Quantum Scattering of a System with Internal Degrees of Freedom}

\author{A.~H. Castro Neto$^{1,2}$ an A.O.~Caldeira$^{2,3}$}

\bigskip

\address{$^{1}$
Institute for Theoretical Physics\\
University of California\\
Santa Barbara, CA, 93106-4030}

\smallskip

\address{$^{2}$
Department of Physics, University of Illinois at Urbana-Champaign\\
1110 West Green Street, Urbana, Illinois, 61801-3080, U.S.A.}

\smallskip

\address{$^{3}$
Instituto de F\'isica ``Gleb Wataghin"\\
Universidade Estadual de Campinas\\
13083-970 Campinas, SP , Brazil.}

\maketitle

\begin{abstract}
We study the quantum tunnelling of a very complex object of which only
part is coupled to an external potential ( the potential barrier ). We
treat this problem as the tunnelling of a particle (part of the system
affected by the potential) coupled to a ``heat bath" ( all of the remaining
degrees of freedom ) from the point of view of the wavefunction of the whole
system. We obtain an effective scattering equation for this wavefunction
which resembles the integral equation for the wavefunction of the particle in
the absence of the bath. We show that the effect of the interaction is to
renormalize the mass and the potential of the problem and that these new
parameters can be solely described in terms of the spectral function of the
excitations of the system.
\end{abstract}
\bigskip
\renewcommand{\baselinestretch}{1.656}   


\narrowtext

The problem of the quantum mechanical tunnelling of non-isolated particles
has been attracting a great deal of attention for over one decade now. People
have been maily interested in the so-called dissipative quantum tunnelling
which turns out to be of crucial importance to problems involving either
macroscopic quantum tunnelling \cite{caldeira2} or macroscopic quantum
coherence \cite{leggett} .Although the mathematical details can vary from
one problem to the other, the basic idea behind all of them is always the
same. One tries to mimic the influence of a very complicated set of degrees
of freedom by a set of non-interacting harmonic modes with an appropriate
spectral function previously chosen \cite{caldeira2,leggett,caldeira1,feynman}.
The next step is to trace out these harmonic modes from the dynamical or
equilibrium density operator of the whole system, depending on the problem one
wants to attack, and use the so-called reduced density operator of the system
in order to obtain the desired information.

The problem we want to face in this work is not of the same kind as the
above-mentioned ones. What we are now interested in is the quantum tunnelling
of a very complex system, say, a huge molecule or a very heavy nucleus, of
which only a small set of variables can couple to an external potential.
We prepare our complex system as a wave packet moving towards that barrier
and ask the following question:``How do the extra modes ( the ones not
feeling the potential barrier ) affect the transmission coefficient of the
complex system as a whole?"

As an example, suppose we want to study the tunnelling of a very complex
nucleus through a potential barrier of purely electromagnetic origin. In
this case only the protons will feel the presence of the potential barrier
although they will still be coupled to the neutrons via strong interaction.
If there was no external potential, the total momentum of the system would be
conserved. However, the presence of this barrier breaks the translation
invariance of the problem and the nucleus will be either reflected or
transmitted. This takes place through the motion of the protons which on their
turn influence the motion of the neutrons. Therefore, the latter will
ultimately affect the motion of the center of mass of the system and we can
test it by studying the effective reflection and transmission coefficients
of the nucleus as a whole.

In order to simplify matters and to keep away from intricate microscopic
models we shall model our complex system as a particle (the relevant degrees
of freedom) coupled to a bath ( the complex set of internal excitations
of the system) subject to a potential barrier acting only on the particle
``coordinate".The bath on its turn will be described by a set of
non-interacting
oscillators and due to our hypothesis of its high complexity we shall also
define its spectral function and ``internal temperature".

Therefore, from now on, we shall be employing the same terminology as for the
general problem of quantum dissipation bearing in mind that it should be taken
seriously only in a very strict sense. We shall emphasize
this remark whenever it is necessary.

Although stated in this way this problem resembles a lot the one of the
transmission of a dissipative particle \cite{bruinsma} they are not quite
the same. For the latter the boundary condition must be entirely different
from the one we shall be using in the present work. Here, the ``system-plus-
reservoir" move together in the same state of motion towards the barrier. It
is clear that the presence of the external potential may cause ``internal
dissipation" due to the flow of energy from the particle to the bath.
Nevertheless, the total energy of the system will be conserved.

It will be shown that the system tunnels as a particle with renormalized mass
through a temperature dependent effective potential barrier. Moreover, the
parameters of this effective scattering theory will be dependent only
on the form of the spectral function of the internal excitations of the
system.

Our approach in this article is quite different from the usual one
for dealing with the quantum dynamics of non-isolated systems. Instead of
treating the time dependent problem using path integrals we will
use a time
independent approach which resembles the problem of scattering
by an external potential, in other words, we will write
the time independent Schr\"odinger equation for $N+1$ bodies
in the approximation where the internal degrees of freedom
act like a heat bath for the particle that feels the external potential.

Our starting point will be the well-known hamiltonian
of a particle
coupled linearly to $N$ harmonic oscillators
which is described as follows \cite{caldeira1,Hakim},
\begin{equation}
H=\frac{p^2}{2 m} + V(x) - \sum_{i=1}^{N} C_{i} x_{i} x
+\sum_{i=1}^{N} \left(\frac{p_{i}^2}{2 m_{i}} + \frac{m_i \omega^2_i
x^2_i}{2} + \frac{C^2_i x^2}{2 m_i \omega^2_i}\right)
\end{equation}
where $p$ and $x$, $p_i$ and $x_i$, are the momentum and the coordinate
of the particle and the oscillators, respectively, $m$ is the ``bare"
mass of the particle, $\omega_i$ and $m_i$ are the frequency and the
mass of the $i^{th}$ oscillator, respectively, $C_i$ is the coupling
constant between the particle and the oscillators and $V(x)$ is the
external potential applied to the particle.

It was shown by Hakim and Ambegaokar \cite{Hakim} that the hamiltonian (1)
can be diagonalized by a unitary transformation
when there is no external potential ($V = 0$). In this case we obtain
a hamiltonian where the center of mass of the whole system, particle
plus oscillators, is decoupled from a new set of noninteracting
oscillators. The diagonalized hamiltonian can be written
as \cite{Hakim},
\begin{equation}
H'= U^{\dag} H U = \frac{P^2}{2 M} + \sum_{j} \hbar \Omega_j
a^{\dag}_j a_j
\end{equation}
where $U$ is a unitary transformation, $P$ is the momentum of
the center of mass of the system and $M$ is the renormalized mass
of the system which is given by,
\begin{equation}
M = m + \sum_{j=1}^{N} \mu_j
\end{equation}
with
\begin{eqnarray}
\mu_j=\frac{C^2_j}{m_j \omega^4_j}.
\nonumber
\end{eqnarray}

The operators $a^{\dag}_j$ and $a_j$ are the
creation and annihilation operators for the new harmonic modes
with frequency $\Omega_j$ which are solutions of the
equation
\begin{eqnarray}
\sum_{i=1}^{N} \frac{C^2_i}{m m_i \omega^2_i (\Omega^2_j-\omega^2_i)} = 1.
\nonumber
\end{eqnarray}

The connection with the dissipative problem can be made if we introduce
the spectral function of the oscillators,
\begin{equation}
J(\omega) = \sum_{i=1}^{N} \frac{ \pi C^2_i}{2 m_i \omega_i} \delta(\omega
-\omega_i)
\end{equation}
which in the ohmic regime is described by\cite{caldeira1},
\begin{eqnarray}
J(\omega)= \eta \omega \Theta(\omega_c-\omega)
\nonumber
\end{eqnarray}
where $\eta$ is the damping coefficient and $\omega_c$ is the
cut-off frequency of the bath. However, one should notice that this
kind of assumption for the behavior of the spectral function is not
appropriate for the present problem.

Observe that in infinite systems the total mass $M$,
eq.~(3), may diverge depending on the behavior of (4) at low frequencies.
This shall not be the case for the sort of systems in which we are
interested in this paper. We shall restrict ourselves to finite systems
and these are bound to present a gap in their spectrum of excitations.
In other words, the spectral function (4) will always have a natural low
frequency
cutoff.

In order to proceed further we include the external potential in the
problem. Following the route established in ref.~\cite{Hakim}
we include a new term in (2) which is given by $U^{\dag} V U$ where
\begin{equation}
U = e^{-p \sum_j K_j(a^{\dag}_j - a_j)}
\end{equation}
with
\begin{equation}
K_j = \sum_{l=1}^{N} \frac{\mu_l \Omega_j^{1/2}}{(2 \hbar)^{1/2}
M \left(\Omega^2_j - \omega^2_l\right)}\left(\sum_{i=1}^{N} \frac{\mu_i
\omega^2_i}{\left(\Omega^2_j-\omega^2_i\right)^2}\right)^{-1/2}
\end{equation}
or in terms of the coordinates of the new oscillators we have
\begin{equation}
U^{\dag} \,V \,U= V\left(X-\sum_{j=1}^{N} \sqrt{2 \hbar \mu_j \Omega_j}
\, K_j \, x_j\right),
\end{equation}
where $X$ is the position of the center of mass of the system.

Equations (6) and (7) show that the center of mass is coupled to the
renormalized oscillators via the external potential.
Suppose now that the potential has a finite range, in other words, it
is a potential barrier. Moreover,
since there are no modes of very low  frequencies
in the problem, there will always be a region for
$X$ sufficiently far from the origin where
(7) vanishes. Therefore, outside the scattering region
we have a free particle with mass $M$ and momentum $\hbar q$
and a set of decoupled harmonic oscillators in states described
by occupation numbers $\{N_j\}$. This is a scenario for
scattering since we already know the asymptotic states of
the problem. This is not revealed in our original hamiltonian
(1).

The most natural way to solve this problem is to rewrite the
time independent
Schr\"odinger equation in the representation of the momentum of
the particle
and the Fock space of the oscillators. Adding (7) to
(2) and writing the resulting equation in this representation together
with the above-mentioned incoming condition,
we easily get
\begin{equation}
\Psi_E(k,\{n_j\}) = \delta(k-q) \prod_{j=1}^{N} \delta_{n_j,N_j} -
\sum_{k',\{n'_i\}} \frac{ G(k-k')_{\{n_j\},\{n'_i\}} V(k-k')
\Psi_E(k',\{n'_i\})}{\frac{\hbar^2}{2 M} (k^2-q^2) + \sum_j \hbar \Omega_j
(n_j-N_j)}
\end{equation}
where $\Psi_E$ is the wave function of $N+1$ bodies with total
energy $E = \frac{\hbar^2 q^2}{2 M} + \sum_j \hbar \Omega_j N_j$,
$V(k)$ is the Fourier transform of the potential and finally,
\begin{equation}
G(k-k')_{\{n_j\},\{n'_i\}} = < \{n_j\}| U^{\dag}(k) U(k')|\{n'_i\} >
\end{equation}
is the thermal part of the kernel of equation (8) where $U$ is defined
in (5). Observe that
the effect of the external potential is completely separated
from the coupling to the oscillators which is present only
in (9). This remarkable fact will allow us to study the
problem independently of the form of the external potential.

Notice that when the interaction in turned off, that is, $C_i=0$,
we have $U = I$ where $I$ is the identity operator. Therefore,
$G(k-k')_{\{n_i\},\{n'_j\}} = \delta_{\{n_i\},\{n'_j\}}$ and the solution
is given by
\begin{equation}
\Psi_E(k,\{n_i\}) = \psi^B_{q}(k) \delta_{\{n_i\},\{N_j\}}
\end{equation}
where $\psi^B$ obeys the ``bare" scattering equation
\begin{equation}
\psi^B_{q}(k) = \delta(q-k) - \frac{1}{\frac{\hbar^2}{2 m} (k^2-q^2)}
\sum_{k'} V(k-k') \psi^B_{q}(k')
\end{equation}
as expected.

The solution of eq.~(8) can be very difficult depending on the
shape of the external potential (as in the problem without dissipation,
eq.~(11)). The presence of the kernel (9) creates new problems since
it contains all induced transitions between the states of the oscillators
due to the recoil of the core particle. We would like to
mention that there is a formal solution for (8) for the case
of local interactions ($V(x)=\alpha \delta(x)$) but
the solution is not illuminating due to its complexity

Our approach here is to suppose that there is a huge number of normal modes
and they are in thermal equilibrium. Notice that we are by no means planning
to model the internal modes of our complex system as an infinite reservoir.
By thermal equilibrium we only mean a fixed average energy which allows us
to define a parameter, the ``internal temperature", we shall loosely refer
to as simply temperature. What we aim at with this approximation is to have
the system of oscillators being only very weakly influenced by the recoil
of the particle. Although this hypothesis is more reasonable for high
temperatures we shall assume it is also the case at very low temperatures.
We then replace (9) by its average value in thermal equilibrium
which is given by,
\begin{equation}
G(k)_{\{n_i\},\{n_j\}}= \delta_{\{n_i\},\{n_j\}} <G(k)>_T
\end{equation}
where
\begin{equation}
<G(k)>_T =\frac{ tr_R \left(e^{-\beta H_R} G(k)\right)}
{tr_R \left(e^{-\beta H_R}\right)}
\end{equation}
where $tr_R$ means the trace over the reservoir coordinates,
$H_R$ is the bath hamiltonian given by the last term on the
r.h.s. of (2) and $\beta = (K_B T)^{-1}$ where $T$ is temperature
($K_B$ is the Boltzmann constant).

If we substitute (12) in (8) we find the effective scattering
equation for the center of mass,
\begin{equation}
\psi_{q}(k) = \delta(k-q) - \frac{1}{\frac{\hbar^2}{2 M}(k^2-q^2)}
\sum_{k'} V^{eff}(k-k',T) \, \psi_q(k')
\end{equation}
where
\begin{equation}
V^{eff}(k,T) = V(k) <G(k)>_T
\end{equation}
is the effective potential felt by a ``dressed" particle with mass
$M$. Now the kernel of the scattering equation
depends on the temperature and the coupling.
The solution of (14) will provide us with
the reflection and transmission coefficients for the scattering of
the particle in the presence of a large set of internal degrees of
freedom.
It is worth mentioning that the wave function $\psi$ {\it is not}
the wave function for the bare particle but {\it it is} the effective
wave function for the particle dragging the surroundings in its motion.

We can evaluate the thermal part of
the kernel given in (13). We can show, after some
straightforward, but lengthy, algebra that
\begin{equation}
<G(k)>_T = e^{-\Gamma(T) k^2}
\end{equation}
where
\begin{equation}
\Gamma(T)=\frac{1}{2} \sum_j K_j^2 \left(1 + 2 \, \overline{n}_j\right),
\end{equation}
and
\begin{eqnarray}
\overline{n}_j = \frac{1}{e^{\beta \hbar \Omega_j}-1}
\nonumber
\end{eqnarray}
is the Bose occupation number.

If we substitute (16) in (15) and transform the problem to  real
space we find that the effective potential is written as
\begin{equation}
V^{eff}(X,T) = \int dX' \, \, V(X') \, \, \Delta(X-X',T)
\end{equation}
where
\begin{equation}
\Delta(X,T) = \frac{1}{\sqrt{4 \pi \Gamma(T)}} e^{-\frac{X^2}{4
\Gamma(T)}}
\end{equation}
and therefore if the interaction is turned off ($\Gamma \to 0$)
we obtain the result that the effective potential is exactly
the bare potential ($Lim_{\Gamma \to 0}\Delta(X,T) = \delta(x)$).

First, we note that the area under the potential is conserved,
\begin{equation}
S=\int dx  \, V^{eff}(x) = \int dx \, V(x)
\end{equation}
as we can easily check. Moreover, it is easy to show  that
the height of the potential is a monotonic decreasing function
of $\Gamma(T)$. Since the area is conserved we conclude that
the width of the potential must increase.

As an illustration of our results consider the simple case of a gaussian
potential,
\begin{equation}
V(x) = V_0 \exp\left\{- \frac{x^2}{a^2}\right\}.
\end{equation}
It is very easy to obtain from (18) that the effective potential
is given by,
\begin{equation}
V^{eff}(x) = \frac{V_0}{\sqrt{1 + 4 \frac{\Gamma^2}{a^2}}}
\exp\left\{-  \left(1 + 4 \frac{\Gamma^2}{a^2}\right)^{-1}
\frac{x^2}{a^2}\right\}.
\end{equation}
Observe that the effect of interaction was to decrease
the height of the potential and increase its width accordingly
 (see equation (20)). Indeed we have,
\begin{equation}
\frac{V(0)}{V^{eff}(0)} = \frac{<x>_{eff}}{a} =
\sqrt{1 + 4 \frac{\Gamma^2}{a^2}}.
\end{equation}

The effect on the transmission coefficient is readily obtained.
 From energy conservation we have,
\begin{eqnarray}
\hbar q &=& \sqrt{2 M (E-E_T)};
\nonumber
\\
\hbar\kappa &=& \sqrt{2 M (V_{eff}(0) - E +E_T)},
\end{eqnarray}
where $E$ is the total energy of the system whereas
$E_T = \sum_j\hbar \Omega_j \overline{N}_j$ stands for its internal energy.
 For temperatures such that  $\kappa a >> 1$, we can show
that the transmission is given by,
\begin{equation}
|t|^2 \simeq 16 \left(\frac{k \kappa}{k^2 + \kappa^2}\right)^2
e^{-4 \kappa <x>_{eff}}.
\end{equation}

The exponent of the transmission coefficient in (25) can be easily analyzed
if we write it as
\begin{eqnarray}
4\kappa <x>_{eff}= 4\sqrt{2 M V_0 a^2} \sqrt{\sqrt{1+ 4 r^2} - \left(
1+ 4 r^2 \right) f }
\end{eqnarray}
where $r = \Gamma/a$ , $ f = E_{K} / V_{0}$ and $ E_K = E - E_T $
 which is the energy of the center of mass of the system.
Observe that we have two effects to take into account in (26). The increase
of the effective mass, $ M $, of the particle and its dependence on the
parameter $r$.

The mass dependence has an obvious influence on the transmission coefficient
because (26) monotonically increases with it. Therefore the transmission
will steadly decrease as a function of the effective mass which is an expected
result.

On the other hand, the dependence of the transmission coefficient on the
combination of the effective height and width of the potential is not so
simple. For a given spectral function one can compute the minimum effective
range of the potential by using (17) for $ T=0 $.This provides us with the
minimum value of $r$ since $\Gamma$ is an increasing function of temperature.
Now, it is worth noticing that (26) has a maximum at $ 1/4f$ and
vanishes at $ 1/2f$ when $ f << 1 $ . Therefore one can say that
as the coupling is switched on there is a reduction of $|t|^2$ as compared to
its non-interacting value. Furthermore,
as one increases the temperature there is
first a reduction of the transmission coefficient before it starts to increase
to its maximum value. Actually we are assuming that in the limit of small $f$
expression (25) is a good approximation for any value of
$ r_{min}< r < 1/2f$. Actually, this is true because in this
range $ \kappa a > 1$.

One can therefore conclude that the effect of the coupling to the
internal degrees of freedom of the system tends to reduce its rate of
transmission through potentials barriers. Actually there is a competition
between this effect and the one related to the decrease of the potential
barrier seen by the system as its center of mass kinetic energy is increased.
The latter wins when $1/4f< r< 1/2f$ and
this causes the transmission coefficient to increase in this region of energy.

This result is in agreement with those obtained in the study of dissipative
quantum tunnelling and coherence \cite{caldeira2,leggett} which say that
the coupling of the form (1) to a heat bath inhibts quantum mechanical
effects. Actually at this point one should be really tempted to draw some
conclusions about dissipative quantum transmission from the results we have
got in this paper. However, although we think that the final results for the
latter  will not differ much from the present ones,
the reader should be warned that the boundary
conditions appropriate to that problem are completely different from the one
we used here and,in fact, it is not even clear to us whether model (1) in its
present form should be the one to describe this phenomenon. The application
of our present results with a spectral function $ J\left(\omega \right) =
\alpha\omega^n $ should be very carefully interpreted since we are treating
only finite $environments$ in this work. It would only make sense for very
large $n$ since in this way we should be trying to simulate a gap in the
spectrum of excitations of the system.

In conclusion, we have derived an effective scattering equation for
a particle coupled to a large, but finite, set of oscillators
(internal degrees of freedom) which resembles the scattering
equation for an isolated particle. We found that the main effects
in the scattering problem are the increase of the mass due to the
dragging of the oscillators and a change in
the scattering potential which depends on the
temperature and $ J\left(\omega\right)
$(through the coupling constants of the problem).
We showed that the results for
high and thin barriers can be obtained independently of the specific form of
the
bare external potential and they imply in the reduction of the transmission
coefficient for tunnelling at low center of mass energies.
 The main assumption in our
derivation is that the oscillators retain
their thermal equilibrium during the scattering process.

A.~H.~Castro Neto gratefully acknowledges M.~P.~ Gelfand
and A.~J.~Leggett for useful comments, F.~Guinea for illuminating
discussions, Conselho Nacional de Desenvolvimento Cient\'ifico
e Tecnol\'ogico, CNPq (Brazil), for a scholarship.
A.~O.~Caldeira also wishes to acknowledge the support from
the Conselho Nacional de Desenvolvimento Cient\'ifico
e Tecnol\'ogico, CNPq (Brazil).
This work was partially supported by
the National Science Foundation through grants no. PHY 89-04035 and
DMR 92-14236.

\end{document}